\documentstyle[preprint,aps]{revtex}

\newcommand{\be}{\begin{equation}}
\newcommand{\ee}{\end{equation}}
\newcommand{\lsgtag}{{\rm LSG}}
\newcommand{\adtag}{{\rm Ad}}
\newcommand{\omegabohr}{{\omega_{\rm B}}}
\newcommand{\mubohr}{{\mu_{\rm B}}}
\newcommand{\vfermi}{{v_{\rm F}}}

\def\draftfig{1}

\ifnum\draftfig=0
  \input epsf.sty
\fi

\begin{document}

\draft

\preprint{}


\title{Observing the Berry phase in diffusive conductors: \\
	Necessary conditions for adiabaticity}

\author{Daniel Loss}

\address{Department of Physics and Astronomy, University of Basel,\\
Klingelbergstrasse 82, 4056 Basel, Switzerland}

\author{Herbert Schoeller}

\address{Institut f{\"u}r Theoretische Festk{\"o}rperphysik,
Universit{\"a}t Karlsruhe,\\
Engesserstrasse 7, Postfach 6980, 7500 Karlsruhe, Germany}

\author{Paul M. Goldbart}

\address{Department of Physics, University of Illinois at Urbana-Champaign,\\
1110 West Green Street, Urbana, Illinois 61801-3080, USA}

\date{\today}

\maketitle

\begin{abstract}

In a recent preprint (cond-mat/9803170), van~Langen, Knops, Paasschens
and Beenakker attempt to re-analyze the proposal of Loss, Schoeller and
Goldbart (LSG) [Phys.\ Rev.\ B~{\bf 48\/}, 15218 (1993)] concerning Berry phase
effects in the magnetoconductance of diffusive systems. Van Langen et
al.\ claim that the adiabatic approximation for the Cooperon previously
derived by LSG is not valid in the adiabatic regime identified
by LSG.  
It is shown that the claim of van~Langen et al.\ is
not correct, and that, on the contrary, the magnetoconductance does exhibit
the Berry phase effect within the LSG regime of adiabaticity.
The
conclusion reached by van~Langen et al.\ is based on a misinterpretation
of field-induced dephasing effects, which can mask the Berry phase (and
any other phase coherent phenomena) for certain parameter values.
\end{abstract}



\section{Introduction}
The Berry phase~\cite{berry} remains a fascinating subject with many
consequences
in a variety of physical systems~\cite{review}.
Some time ago we proposed~\cite{LGB1,LGB2,LGlong,LGexp,LSG} a number of
scenarios in condensed matter settings 
where the Berry phase manifests itself
in the phase-coherent quantum dynamics
of a particle carrying a  spin
and moving through orientationally inhomogeneous magnetic fields
${\bf B}({\bf x})$. Such manifestations of the Berry phase can occur, 
e.g., in semiconductors or metals in the form of 
persistent currents~\cite{LGB1,LGB2,LGlong,LGexp} or 
oscillations of the magnetoconductance or universal
conductance
fluctuations~\cite{LGB2,LSG}. As recognized early on\cite{LGB2}, all
these effects share the common
feature that the orbital motion of the particle is modified by 
the Berry phase in very much the same way as it is in 
well-known phase-coherent phenomena based on the Aharonov-Bohm effect.

The first experimental evidence for such a Berry phase effect
has been recently found in semiconductors~\cite{morpurgo}, in 
which a local effective magnetic field is produced
via the Rashba effect.

However, whereas Aharonov-Bohm effects occur regardless of the 
strength $B$ of the field, 
Berry phase effects appear only in the adiabatic limit,
i.e., for sufficiently large magnetic fields. This limit
requires that---roughly speaking---the typical orbital frequency of the particle
carrying the spin through the field is much
smaller than the  precession
frequency of the spin around the local field direction. In this
limit, the spin
will remain in its instantaneous
eigenstate, i.e., will continuously align itself along the {\it local\/}
field direction ${\bf B}({\bf x})$ as it moves through
the magnetic field texture. If, in addition, the particle trajectory is
closed,
the spin will acquire a Berry phase, which is purely geometric in character.
As spin and orbital motion couple via the inhomogeneity of the field,
the Berry phase can ultimately enter the orbital part of
the effective Hamiltonian
in the same way that the Aharonov-Bohm phase does.

There seems to be general agreement that once the adiabatic limit is
reached the results  
found previously~\cite{LGB1,LGB2,LGlong,LGexp,LSG} 
are correct. The central question then is: What is the proper
criterion for the adiabatic regime?  Again, there is no issue 
of contention in ballistic rings, e.g., 
for which adiabaticity is reached when $\omegabohr t_o\gg 1$, where 
$\omegabohr$ is the Bohr frequency (to be defined below), and
$t_o$ is the typical time it takes the particle to go around the ring once.
This situation occurs, e.g., in clean semiconductors.

But what about diffusive systems, such as normal metal rings?
It is this question that we have previously addressed  in 
great detail~\cite{LSG}
and that has been recently reconsidered  by 
van Langen et al.~\cite{langen}, who claim to 
reach a rather pessimistic conclusion
about the observability of the Berry phase effect---in stark contrast 
to our findings~\cite{LSG}.
It is the purpose of the present paper to show that the claim of van Langen
et al.~\cite{langen} is not correct. To this end,  
we first state the problem of adiabaticity in this section again and then 
provide in the following sections
a general discussion on 
the issue of dephasing induced by
inhomogeneous magnetic fields. This discussion is then followed by
explicit examples that
unambiguously demonstrate the observability of Berry phase 
effects in diffusive systems
of immediate experimental interest.

Now, in the context of weak localization physics we have advanced 
detailed physical and technical arguments~\cite{LSG}
that adiabaticity is reached more easily
in diffusive than in ballistic systems (all other parameters being equal).
The physical explanation for this is simple: In diffusive motion
around, say, a ring, 
the particle spends on the average much more time in a given region of field
direction than
it would do in purely ballistic motion. Thus, there is more time for the
spin to execute
precessions around a given field direction, and thus the spin will have a
higher probability
of aligning itself along the local field direction than it would in purely
ballistic
motion. Translating this picture into more concrete terms for an electron
diffusing
around a $d$-dimensional ring of circumference $L$ with static random
disorder,
adiabaticity is reached
if the Zeeman energy, $\hbar \omegabohr=g\mubohr B/2$, exceeds
the Thouless energy, $E_{\rm Th}=hD/L^2$.
Here $g$ is the electron g-factor, 
$\mubohr$ is the Bohr magneton, 
$D=\vfermi^2\tau/d$ is the diffusion constant 
with $\vfermi$ being Fermi velocity, 
$\tau=l/\vfermi$ is the elastic mean free time, and 
$l$ is the elastic mean free path. 
More generally, we can also allow
for the case in which the field reorients $f$ times as the particle goes around
once the ring.  Whereas the case of $f=1$ is physically 
realizable~\cite{LGlong}, it seems 
very difficult 
to implement cases with $f>1$ experimentally. 
Still, as the conclusions reached by 
van~Langen et al.~\cite{langen} are 
crucially based on the case $f=5$, we shall include this possibility, and
the criterion for adiabaticity as found in Ref.~\cite{LSG} then reads
\begin{equation}
\omegabohr \tau \gg {f\over d} {l^2\over L^2} \sqrt{1-|{\bf N}|} .
\label{adiabaticity}
\end{equation}
Here, the texture--dependent vector ${\bf N}$ is some average of the 
direction of the magnetic field~\cite{LGlong}.
The factor $\sqrt{1-|{\bf N}|}$ accounts for non-uniformity in the 
direction of the magnetic field, and encodes the fact that the adiabatic 
approximation becomes exact, regardless of $\omegabohr$, in the limit of 
a homogeneous field, for which $|{\bf N}|=1$. 
In the following discussion, however, we shall---for the sake of 
simplicity---omit this factor, noting that its inclusion would render  
the criterion even less stringent)~\cite{footnote0}.
As in metals one typically has 
$\tau$ of the order of $10^{-14}\,{\rm s}$,  
$g=2$ and $l=10^{-8}\,{\rm m}$, 
we should have, for a ring of
circumference $L=10^{-6}\,{\rm m}$, 
magnetic fields at least of the order of $100-1000$~Gauss to be within
the adiabatic regime.  Note that without the diffusive
factor, $(l/L)^2=10^{-4}$, the required fields would be too large to be
attainable experimentally 
(i.e., on the order of $100-1000\,{\rm T}$).

The regime of adiabaticity defined in Eq.~(\ref{adiabaticity}) follows from
a detailed derivation of the Cooperon and Diffuson propagator based on
weak localization techniques  and an adiabatic
approximation scheme~\cite{LSG}.
This adiabatic approximation is performed in the path intgral representation
for the Cooperon (Diffuson). As emphasized in an analogous
discussion of the imaginary-time propagator  in the
context of persistent currents~\cite{LGlong},
the adiabatic approximation can contain additional
angle-dependent terms that are different from the Berry phase, and
these terms can mask the Berry phase in certain physical observables. 
(For an explicit example of such a case, see Sec.~VI~F of Ref.~\cite{LGlong}.) 
The origin of this additonal term can be traced back
to quantum fluctuations of the particle trajectory, which induce
non-smooth variations of the magnetic field (and thereby violate the
\lq\lq smooth variation\rq\rq\ assumption that underlies the adiabatic 
approximation)\cite{LGlong}.
An alternative way to express this point is to say that in certain cases 
the Berry phase can be masked by dephasing effects---in very much the same 
way that the Aharonov-Bohm phase can become unobservable if dephasing 
influences become too large.
Such dephasing effects are difficult to calculate  for a general texture,
but can sometimes be obtained in special
cases for which an exact solution is available
(see Ref.~\cite{LGlong} and below). As suggested in Ref.~\onlinecite{LSG},
it is possible to extend the exact solution for a propagator
containing a single spin-1/2 particle~\cite{LGlong}
to the one containing two spin-1/2 particels.
Indeed, by following this suggestion van~Langen et al.~\cite{langen} re-calculate
the magnetoconductance
for a cylindrically
symmetrical texture, and claim to find 
deviations from our adiabatic solution~\cite{LSG}.
(As we shall show, these deviations are only apparent.)\thinspace\
Van~Langen et al.~conclude 
from this observation that the exact solution does not contain
the Berry phase effect, and thus that the
regime of adiabaticity, given in Eq.~(\ref{adiabaticity}), is 
invalid.
Instead, adopting a suggestion made first by Stern~\cite{stern}, 
van~Langen et al.\cite{langen} claim that it is necessary for the much more stringent 
conditon, 
\begin{equation}
\omegabohr \tau \gg 1 \,\, ,
\label{schrott}
\end{equation}
to be satisfied in diffusive systems before adiabaticity is reached, 
and thus before the Berry phase effect can become observable in the 
magnetoconductance.  It is specifically this claim that is incorrect.  
On the contrary, we will 
show that precisely our adiabaticity criterion, Eq.~(\ref{adiabaticity}), 
is appropriate for diffusive systems, 
and that the observability or
non-observability of the Berry phase crucially
depends on the choice of physical parameters [in the adiabatic
regime given by Eq.~(\ref{adiabaticity})].
Indeed, van~Langen et al.\cite{langen}  concentrate on the rather unphysical choice
that the field winds five times around the ring (i.e., $f= 5$),
and as dephasing effects grow strongly with $f$ (as $f^2$; see below),
it comes as little surprise
that Berry phase oscillations are not discernible in this extreme case.
However, upon choosing $f=1$---the physically most 
relevant case---not only do Berry phase effects  show up in the exact 
solution, but also they agree well with our previously-obtained 
adiabatic predictions.

Van Langen et al.\cite{langen} have studied the issue of adiabaticity also 
in terms of Boltzmann equations.  Due to the coupling of the magnetic
field to the orbital motion of the charged electron
these Boltzmann equations  are valid in the 
diffusive regime defined by $\omega_{\rm c} \tau\ll 1$, where $\omega_{\rm c}$
is the cyclotron frequency. 
As $\omega_{\rm c}$ and $\omegabohr$ are typically of the same order
of magnitude in metals, the regime $\omegabohr \tau \gg 1$ studied by van~Langen et 
al.\ lies {\it outside\/} the physical regime to which their
Boltzmann equations can  legitimately be applied.
Still, even if we adopt their academic point of view
and ignore such orbital effects (i.e. set the electron charge to zero), 
the regime $\omegabohr \tau \gg 1$
is problematic for an additional reason~\cite{LSG}.
If $\omegabohr \tau \gg 1$, the Zeeman rate $\omegabohr$ is large compared to
the elastic collision rate $1/\tau$. In this case
we expect the Zeeman interaction to have
a strong dephasing influence on the orbital motion (for inhomogeneous fields),
especially when $f\gg 1$, and 
the system lies outside the
semiclassical regime in the sense of weak localization
theory (see, e.g., Secs.~4 and 10 of Ref.~\onlinecite{Schmid} 
and below).
This issue has not been
discussed by van~Langen et al.\ in the context of their Boltzmann equations.

Finally, none of the effects discussed by van~Langen
et al.\ in terms of their Boltzmann equation have been shown explicitly
to be related to
the Berry phase. Without such information at hand
it is not possible to tell whether the effects they find are
of dynamical (non-phase-coherent)
or geometrical (Berry phase) origin, as both of them can
occur in an adiabatic approximation to the quantum dynamics.
As we are interested in the Berry phase effect associated
with phase-coherence and occurring
in physical observables, 
we shall not comment any further
on the Boltzmann equation approach of van~Langen et al., and 
instead shall concentrate on the magnetoconductance
expressed in terms of the Cooperon propagator~\cite{LSG}.

Still, we do wish to point out that although we do not agree with
the final conclusions reached by van~Langen et al., we have found 
their work stimulating, inasmuch as it has motivated us to clarify
the issue of dephasing which, in turn, has allowed us to establish more 
concrete predictions about the range of observability of the Berry 
phase in the magnetoconductance of diffusive metallic systems.

\section{Berry Phase and Magnetoconductance}

\subsection{Exact Solution and Adiabatic Approximation}
\label{exactsolution}

We consider a quasi-one-dimensional ring 
of circumference $L$, 
embedded in a magnetic field texture
given by 
${\bf B}=B{\bf n}=B
(\sin{\eta}\cos{2\pi f x\over L},
\sin{\eta}\sin{2\pi f x\over L},
\cos{\eta})$, 
where $x$ is the location on the ring, 
$\eta$ is the tilt angle of the magnetic field, 
and $f(=1,2,3,\ldots)$ is the winding of the magnetic field along the 
propagation direction.
The magnitude $B$ and, in particular, the tilt angle $\eta$
are assumed to be constant. It is this special case that
can be solved exactly (as pointed out in Ref.~\onlinecite{LSG})
along the same lines as discussed in Ref.~\onlinecite{LGlong} for a 
single-spin propagator.  Van~Langen et al.~\cite{langen} were the first 
to write down this solution explicitly for a two-spin propagator.

The magnetoconductance resulting from weak localization corrections
and in the presence of the field texture ${\bf B}$
has been derived in Ref.~\cite{LSG} and reads,
\begin{equation}
\delta g= -{e^2 \over \pi \hbar }{L \over (2 \pi)^2}\sum_{\alpha,\beta=\pm 1}
\langle x,\alpha,\beta|{1\over{\gamma} -{ h}}|x,\beta,\alpha\rangle
\label{mosc}
\end{equation}
where  the effective (non-hermitian) Hamiltonian $h$ is given by
\begin{equation}
h={L^2\over (2\pi)^2}{\partial^2\over \partial x^2} + i
\kappa\, {\bf n}\cdot ({\bf \sigma}_1-{\bf \sigma}_2),
\label{ham}
\end{equation}
where ${\bf \sigma}_i$ (with $i=1,2$) are spin-1/2 Pauli matrices, 
and where
\begin{equation}
\kappa= {\omegabohr \over D} {L^2\over (2\pi)^2}=\omegabohr \tau d
{L^2\over (2\pi l)^2}
\end{equation}
is the
dimensionless adiabaticity parameter [see Eq.~(\ref{adiabaticity})].
The factor $\gamma = (L/2\pi L_{\phi})^2$ is a  damping
constant expressed in terms of the dephasing length $L_\phi$
(which is specified
in more detail below). Note that $\gamma$ is introduced here in a
phenomenological way with the particular {\it ad hoc\/} choice that it be
a c-number and diagonal in spin space.

We now evaluate $\delta g$ explicitly, but instead of using
the exact eigenstates, as was done by van~Langen et al.~\cite{langen}, 
we use an alternative approach
in terms of unitary gauge transformations, 
which has the virtue of making the emergence of the Berry
phase immediately transparent. For this purpose we define
unitary transformations $U$ and $V$ of the form
\begin{equation}
U=V\, e^{{i \pi f\over L} x (\sigma_{1z}+\sigma_{2z})}\,\,  ,
\qquad
V=e^{{i\over 2}\eta (\sigma_{1y}+\sigma_{2y})}\,\, ,
\label{unitary}
\end{equation}
with the property that 
\begin{equation}
{\bf n}\cdot ({\bf \sigma}_1-{\bf \sigma}_2)=U^\dagger\,
(\sigma_{1z}-\sigma_{2z})\,U.
\end{equation}
By noting that 
$U(-i\partial/\partial x)U^\dagger=-i\partial/\partial x-
iU\partial U^\dagger/\partial x$, we find 
\begin{equation}
UhU^\dagger=-(- i{L\over 2\pi}{\partial\over \partial x} -{f\over 2}
[(\sigma_{1z}+\sigma_{2z}) \cos\eta -(\sigma_{1x}+\sigma_{2x})\sin\eta])^2
+i\kappa\, (\sigma_{1z}-\sigma_{2z}).
\label{uhu}
\end{equation}
Next, we  rewrite the matrix elements occurring in $\delta g$:
\begin{equation}
\langle x,\alpha,\beta|U^\dagger{1\over {\gamma} -UhU^\dagger }U
|x,\beta,\alpha\rangle
=\langle x,\alpha,\beta|V^\dagger{1\over {\gamma} -h_{\alpha \beta}}\Pi_{12}V
|x,\alpha,\beta\rangle,
\label{tralala}
\end{equation}
where $h_{\alpha \beta}=UhU^\dagger(-i{L\over 2\pi}{ \partial \over\partial x }
\rightarrow -i{L\over 2\pi}{ \partial
\over\partial x} +{f\over 2}(\alpha+\beta))$ and $\Pi_{12}|\alpha\beta\rangle=
|\beta\alpha\rangle$.
The effective Hamiltonian $h_{\alpha \beta}$ is now diagonal in the 
angular momentum
eigenstates $\langle x| j\rangle =e^{i{2\pi\over L} x j}/\sqrt{L}$, with
$j=0,\pm 1,\pm 2,\ldots$ (imposing periodic boundary conditions),
and we find,
\begin{equation}
\delta g= -{e^2 \over \pi \hbar }{1 \over (2 \pi)^2}\sum_j
\mbox{Tr}_{12}{1\over {\gamma} -{ h(j)}}\Pi_{12}\,\, ,
\label{mosc2}
\end{equation}
where $\mbox{Tr}_{12}$ is the trace in spin space and
\begin{eqnarray}
h(j)&=&-( j -{f\over 2}
(\sigma_{1z}+\sigma_{2z}) \cos\eta)^2 - {f^2\over 2}
(1+\sigma_{1x}\sigma_{2x})\,\sin^2{\eta} \\ \nonumber
&&\qquad
-jf(\sigma_{1x}+\sigma_{2x})\,\sin\eta
+{f^2\over 4} \,(\sigma_{1x}\sigma_{2z}+\sigma_{2x}\sigma_{1z})\,\sin{2\eta}
+i\kappa\, (\sigma_{1z}-\sigma_{2z}).
\label{hfinal}
\end{eqnarray}
Here, we have absorbed the integer $f(\alpha+\beta)/2$ into $j$. Note that two
of the  
eigenvalues of ${f\over 2} (\sigma_{1z}+\sigma_{2z}) \cos\eta$
are given by the (geometric)
Berry phase  $\pm \Phi^g=\pm f\cos\eta$ for an effectively integral spin
\cite{footnote1}.
The term ${f^2\over 2} (1+\sigma_{1x}\sigma_{2x}) \sin^2{\eta}$
provides a source of dephasing
that can mask the Berry phase--and more generally the
Aharonov-Bohm effect (see, Sec. \ref{dephasing} below).
All the other off-diagonal terms turn out to be irrelevant in the
adiabatic limit (see Sec.~\ref{physical}).
To proceed, we express the above operators
in the $\sigma_z$-basis $\{|1,1\rangle,|1,-1\rangle, |-1,1\rangle,
|-1,-1\rangle \}$.
The Hamiltonian $ h(j)$ then has matrix elements
\begin{eqnarray}
&\langle& \alpha',\beta' | h(j) |\alpha,\beta\rangle = 
\\ \nonumber
\\ \nonumber
&-&\pmatrix{
(j-f\cos\eta)^2 + a  &   jf\sin\eta- b
& jf\sin\eta- b       &   a   \cr
jf\sin\eta- b   &  j^2 +a -i 2\kappa
 &  a    &   jf\sin\eta + b     \cr
jf\sin\eta- b     &   a
& j^2 +a +i 2\kappa
&  jf\sin\eta + b    \cr
a   &    jf\sin\eta + b
&    jf\sin\eta + b
 & (j+f\cos\eta)^2  +a \cr
 }.
\label{matrixh}
\end{eqnarray}
where $a={f^2\over 2}\sin^2{\eta}$, and $b={f^2\over 4}\sin{2\eta}$.
Finding the inverse of ${\gamma} - h(j)$ is then straightforward, and we
finally obtain for the magnetoconductance
\begin{eqnarray}
\label{moscexact}
&\delta g &= -{e^2 \over \pi \hbar }{1 \over 2 \pi^2}\sum_{j=-\infty}^{+\infty}
\{(\gamma +m^2 +f^2)(\gamma +m^2 )^2 +4\kappa^2 (\gamma +m^2 +
f^2\cos^2{\eta} +{f^2\over 2} \sin^2{\eta}) \}  \\ \nonumber
&\times& \{ [\gamma+(m-f)^2]  [\gamma+(m+f)^2]\, (\gamma+m^2)^2  \\ \nonumber
&+&
4\kappa^2
\{[\gamma+(m-f\cos\eta)^2][\gamma+(m+f\cos\eta)^2]+f^2\sin^2{\eta}
(\gamma +m^2 +f^2\cos^2{\eta})\} \}^{-1}\,\, ,
\end{eqnarray}
where $m=j-\Phi$, i.e., we have allowed for an  Aharonov-Bohm
flux $\Phi=2\phi/\phi_0$, with $\phi_0=h/e$ being the flux quantum.
The foregoing result is exact and is seen to be identical
to the one obtained  by van~Langen et al.~\cite{langen} (for their choice
$d=2$).
However, our alternative derivation
has led us to a form in which the Berry phase contribution is made manifest
in the terms of the form $(m\pm f\cos\eta)^2$.

Next we go over to the adiabatic limit, defined here by
$\kappa\gg 1/ (2\pi)^2$, which,
for $f=1$, is equivalent to $\omegabohr\tau\gg l^2/(L^2d)$ 
[see Eq.~(\ref{adiabaticity})].
(Below, in Sec.~\ref{observability}, we give explicit numerical
values of $\kappa$ for which adiabaticity is reached.)\thinspace\ 
In this limit we may drop the terms independent of $\kappa$ in Eq.~(\ref{moscexact})
(this is justified as terms with large $j$ give a negligible contribution
to $\delta g$). Thus, in the adiabatic limit we finally get
\begin{eqnarray}
\label{moscadiabatic}
\delta g^{\adtag} &=& -{e^2 \over \pi \hbar }{1 \over (2 \pi)^2}
\sum_{\alpha=\pm 1}\sum_{j=-\infty}^{+\infty} \\ \nonumber
&\times&{[\gamma +(m +\alpha f\cos{\eta})^2] +(f^2/2)\sin^2{\eta}
\over [\gamma+(m-\alpha f\cos\eta)^2][\gamma+(m+\alpha f\cos\eta)^2]+
(\gamma +m^2 +f^2\cos^2{\eta})f^2\sin^2{\eta}}\,\, ,
\end{eqnarray}
where the sum over $\alpha=\pm 1$ has been introduced artificially
for later convenience. Note that the  Berry phase
$\Phi^g=f\cos\eta$ couples to the momentum like the Aharonov-Bohm
phase does, i.e., via $j-\Phi-\alpha \Phi^g$.
We note that the remaining $\eta$-dependence can not be 
accounted for by this type of coupling to the momentum. 
We particularly emphasize that (apart from the flux appearing in $m=j-\Phi$)
the adiabatic limit of the magnetoconductance $\delta g^{\adtag}$ 
is independent of the field amplitude $B$;
thus, increasing the field
further, say up to $\omegabohr\tau\gg 1$ (cf. Eq. (\ref{schrott})), has no effect.

It is now instructive to compare Eq.~(\ref{moscadiabatic})
with the one previously
derived~\cite{LSG} for arbitrary textures and in the adiabatic approximation
scheme for the Berry phase. The latter result reads 
\begin{eqnarray}
\label{moscLSG}
\delta g^{\lsgtag}&=& -{e^2 \over \pi \hbar }
{L'_\phi \over 2L} \sum_{\alpha=\pm 1}{\sinh{(L/L'_\phi)}\over
\cosh{(L/L'_\phi)} -\cos{(2\pi(\Phi+\alpha f\cos\eta))}}\\ \nonumber
&=& -{e^2 \over \pi \hbar }
{1 \over (2 \pi)^2}\sum_{\alpha}\sum_{j=-\infty}^{+\infty}
{1 \over \gamma'+(m-\alpha f\cos\eta)^2} \\ \nonumber
&=& -{e^2 \over \pi \hbar }
{1 \over (2 \pi)^2}\sum_{\alpha} \sum_{j=-\infty}^{+\infty}
{\gamma' +(m +\alpha f\cos{\eta})^2
\over [\gamma'+(m-\alpha f\cos\eta)^2][\gamma'+(m+\alpha f\cos\eta)^2]}\,\, ,
\end{eqnarray}
where, again, $m=j-\Phi$, and $\gamma'=(L/2\pi L'_{\phi})^2$, and where
we have used some identities
to facilitate comparison. Note that in general $\gamma\neq \gamma'$ (see
below).
The virtue of $\delta g^{\lsgtag}$ is that it is valid
for arbitrary field textures (with the appropriate Berry phase~\cite{LSG}).
It is thus important to understand its relation to the special but
exactly solvable case.

Now, by comparing $\delta g^{\lsgtag}$ with $\delta g^{\adtag}$ we see
that the two expressions
have the same structure with respect to the Berry phase, $\Phi^g=f\cos{\eta}$,
but differ in additional $\eta$- and $f$-dependent terms.
(From now on we put the 
Aharonov-Bohm 
flux $\Phi$ to zero but shall
return to nonzero flux later.)\thinspace\ 
Particularly important
is the additional term in the denominator of $\delta g^{\adtag}$, i.e., 
$f^4\sin^2{\eta}\cos^2{\eta}$ (the physical origin of such additional terms
is discussed below in Sec.\ref{physical}). It is this term
that acts as a {\it dephasing source\/} for certain
tilt angles and windings $f$ by suppressing
the ``resonance peaks" that would occur at
integral values of the Berry phase $\Phi^g=f\cos\eta$ (for small
enough $\gamma'$).
For $f>1$ the suppression due to this term is so strong that
all resonances except the ones at $\eta=0,\pi/2,\pi$ become
masked, i.e., these resonances due to the Berry phase
are no longer visible in graphs of $\delta g^{\adtag}$
versus $\eta$, whereas they do show up in $\delta g^{\lsgtag}$
{\it provided\/} one chooses $\gamma'$ to be independent of the tilt 
angle $\eta$ (and sufficiently small).
It is this  {\it ad hoc\/} choice for $f$ and $\gamma'$ that has
been adopted by van~Langen et al.~\cite{langen}.
In particular, they choose
$f=5$ and  a constant $\gamma=0.4053$. 
As in this case $\delta g^{\lsgtag}$
and $\delta g^{\adtag}$ behave differently for $\gamma=\gamma'$ 
(see Fig.~3 of Ref.~\cite{langen}),
van~Langen et al.\cite{langen} conclude that $\delta g^{\adtag}$
is not showing adiabatic behavior and, thus, that our criterion for
adiabaticity, Eq.~(\ref{adiabaticity}), is not correct.
However,  this conclusion of van~Langen et al.\ is premature. 
There are two main reasons for this: First, they have ignored 
the issue of dephasing induced by the inhomogeneity of the magnetic field, and 
connected to this, second, the issue of the self-consistency of the 
semiclassical approximation on which the derivation of the Cooperon 
propagator rests.  We now discuss these issues in turn, and then 
present physical examples to illustrate the general discussion.

\subsection{Dephasing due to Magnetic Fields}
\label{dephasing}

The {\it ad hoc\/} choice by van~Langen et al.~\cite{langen}
of putting $\gamma=\gamma'$ and choosing
them to be independent of $\eta$ means that
$\delta g^{\adtag}$ and $\delta g^{\lsgtag}$ do
{\it not\/} describe the same physical situation. This is so
for the following reason.
First we note again that the dephasing parameters
$\gamma, \gamma'$ are ``put in by hand" into
the Cooperon to account for dephasing in a phenomenological way (this
is just dictated by the complexity of the involved many-body problem and by
our inability to address this issue in a more systematic way in general).
In the derivation of $\delta g^{\lsgtag}$ dephasing due to the field is
only taken into account {\it a posteriori\/} in terms of a phenomenological
parameter $\gamma'$, while
the exact solution, Eq.~(\ref{moscexact}), not only includes the Berry phase
but simultaneously also those dephasing effects that are caused by the field
through the Zeeman coupling. The remaining dephasing effects in
$\delta g$ or $\delta g^{\adtag}$ are then included via
the phenomenological parameter $\gamma$. Obviously, $\gamma$ and $\gamma'$
are in general different  for the same physical situation.

Next, it is a well-known fact in the context of weak-localization
phenomena~\cite{Aronov}  that
dephasing in general depends on the magnetic field ${\bf B}$
penetrating the sample
(as we must allow for there to be any Zeeman interaction at all).
Most importantly,
$\gamma'$ not only depends, in general, on the
magnitude $B$ of the field but also on its tilt angle $\eta$ that 
the field makes with
the $z$-axis perpendicular to the ring plane. 
(This is already so even without Zeeman terms, see, e.g., Sec.~2 of
Ref.~\onlinecite{Aronov}. There can be little surprise that
the angle dependence becomes even more pronounced in the presence of our
inhomogeneous Zeeman interaction).\thinspace\ 
The various dephasing effects are accounted for phenomenologically 
in terms of
dephasing lengths~\cite{Aronov},
$1/L_\phi^2=1/(L_\phi^0)^2 +1/(L_\phi^B)^2$, where the dephasing
length $L_\phi^0$ contains all field-independent contributions,
such as the one coming from inelastic collisons of the diffusing
electron with, say, phonons,
$L_{\phi}^{in} =\sqrt{D{\tau}_{in}}$,
where the dephasing time $\tau_{in}$ is some inelastic scattering time.
The magnetic length $L_\phi^B$ contains all effects coming
from the field penetrating the sample.

If now $L_\phi^B\ll L$ for some
field configurations, we no longer expect to see phase-coherence in general.
As a matter of fact, in Sec.~IV of Ref.~\onlinecite{LSG} we have estimated
the upper bound of the dephasing length (due to the inhomogeneous 
Zeeman interaction) in metallic films to be given
by the characteristic field-reorientation length 
$l_B=|\nabla ({\bf B}/B)|^{-1}$.
This estimate follows from the observation that quantum corrections begin
to be eliminated when the largest phase-coherent paths enclose roughly
one quantum of Berry flux.
For the symmetric texture considered here
we find $l_B=L/(2\pi f|\sin\eta|)$. Obviously,
for certain tilt-angles and for $f\gg 1$ this upper bound on the
dephasing length quickly becomes smaller than $L$. Translated into a
dephasing parameter $\gamma=(L/2\pi L_\phi^B)^2$, this estimate reads
\begin{equation}
\gamma > f^2 \sin^2{\eta} ,
\label{estimate}
\end{equation}
i.e., we see that the dephasing becomes explicitly $\eta$-dependent 
and grows like $f^2$.

Thus, it is by no means surprising that the exact solution
confirms this general expectation, 
in the sense that explicit dephasing terms are present in 
$\delta g$
that are field-dependent and which can become
so large, for {\it particular\/} field inhomogeneities, that
they completely suppress
the resonances in the magnetoconductance, Eq. (\ref{moscexact}), 
with respect to the Berry phase~\cite{footnote3}, no matter how large $\omega_B$
is.
Of course, as implied by above discussion leading to Eq. (\ref{estimate}),
such a dephasing effect must also be accounted 
for explicitly in $\delta g^{\lsgtag}$, Eq. (\ref{moscLSG}), by an appropriate
choice for the phenomenological damping parameter $\gamma'$.
In particular, in view of the estimate given in Eq.~(\ref{estimate}),
it is reasonable to make the Ansatz
$\gamma'=f^2\sin^2(2\eta)$~\cite{footnote5}.
Then, choosing 
the dephasing parameter of $\delta g^{\adtag}$ to be constant (i.e. $\eta$-independent)
and much smaller than unity, say $\gamma=10^{-2}$,
we see that the qualitative discrepancy between 
$\delta g^{\lsgtag}$ and $\delta g^{\adtag}$ disappears:
Both expressions show no resonances  (away from $\Phi^g=0,1$).
(We note that as $\gamma$ and $\gamma'$ are introduced phenomenologically 
anyway, there is no need to get quantitative agreement,
and it suffices to find the same qualitative suppression  of the
resonances for $f>1$ in both $\delta g^{\lsgtag}$ and $\delta g^{\adtag}$.
We shall not be making any further use of this Ansatz for $\gamma'$.)

The suppression of the Cooperon due to homogeneous fields is
standard~\cite{Aronov}; the discussion above 
shows that additional dephasing is induced by the field
inhomogeneity.
The advantage of having the exact solution for $\delta g$, Eq. (\ref{moscexact}),
at hand
is that we can now 
calculate the field-dependence of such dephasing terms explicitly;
this allows us to make more precise
statements than before~\cite{LSG} about the regime
in which one can expect to observe consequences of the Berry phase
(see Sec.~\ref{observability} below).

\subsection{Self-consistency of the Semiclassical Approximation}
\label{semiclassical}

The  magnetoconductance correction $\delta g$ is expressed in terms
of the Cooperon propagator. The derivation of the Cooperon is, in turn,
performed within the {\it semiclassical limit\/}. In particular, this 
means that ``back-reaction effects", i.e., 
non-phase-coherent dynamical effects of the
field-dependent Zeeman term on the {\it orbital motion\/} are assumed
to be negligibly small throughout.
This is a fundamental assumption in weak-localization theory~\cite{Schmid},
and it was explicitly adopted in our
derivation of the Cooperon and of $\delta g^{\lsgtag}$, too. 
(This is emphasized, e.g., in App.~A of Ref.~\onlinecite{LSG}.)\thinspace\ 
Evidently, dephasing effects such as the ones discussed in the previous subsection
are nothing 
but such back-reaction effects.  Thus, if  dephasing becomes
so large (as turns out to be the case in the adiabatic limit and
for $f>1$) that phase-coherence is completely suppressed in the orbital 
part, the semiclassical approximation breaks down and the self-consistency 
of the entire treatment is lost~\cite{footnote2}. Consequently, the 
expressions for the magnetoconductance are no longer reliable in the case 
of complete dephasing, and no weight should be put on conclusions drawn 
under such circumstances. 
Obviously, semiclassical and adiabatic approximations are interconnected 
issues, in the sense that the semiclassical approximation might
break down in the adiabatic limit and for certain field configurations.
In other words, adiabaticity alone is not a sufficient criterion
for the observability of Berry phase effects, in addition the system must be
in the mesoscopic regime characterized by phase-coherence.

To summarize our conclusions so far,
we have seen that our adiabaticity
criterion, Eq.~(\ref{adiabaticity}),  is sufficient for reaching
the adiabatic limit involving the Berry phase 
[cf.~Eqs.~(\ref{moscadiabatic}) and
(\ref{moscLSG})].
However, the criterion
does not guarantee (and this was never claimed)
that the Berry phase will be observable under all circumstances.
As a matter of fact, it can happen
that the phase-coherence,
which is necessary for observing such quantum phase phenomena, can be destroyed
by a variety of dephasing sources, in particular also
by magnetic fields penetrating the sample.
If  dephasing  becomes so strong in the adiabatic regime
that quantum phase effects of the orbital motion get completely washed out,
the semiclassical approximation  underlying the
derivation of the Cooperon breaks down
and results based on it (such as $\delta g$)
are no longer reliable.

It is precisely
the issues discussed in the last two
subsections
that have not been taken into consideration
by van~Langen et al.~\cite{langen}. In the light of our discussion it should 
now be clear that the only conclusion
one can draw from the observation made by van~Langen et al.\ (namely the
non-observability of Berry phase effects for $f=5$ within our semiclassical
theory) is that
field textures with $f>1$ suppress phase-coherence very efficiently, 
and thus such extreme textures cannot serve as a general test case
for the existence of the Berry phase and the associated adiabaticity
regime---at least not within the semiclassical regime to which
our results, Eqs.~(\ref{moscexact})-(\ref{moscLSG}), are confined.

\subsection{Observability of Berry Phase Effects for $f=1$}
\label{observability}

Up to now we have mainly concentrated on regimes where $f>1$.
Such regimes, however,
are of little experimental interest (quite apart from the difficulty of how to produce
them) since the Berry phase effect would be masked by the strong dephasing effect
of the field.
The situation, however,  is entirely different for the case where the magnetic field
winds only once around the ring, i.e. when $f=1$ (such field textures can be produced
experimentally~\cite{LGlong}).  
Indeed,  we shall  see now that for $f=1$ the dephasing is sufficiently small
and  the Berry phase has observable consequences  within an experimentally
accessible regime.
We shall illustrate this with two
specific examples: First we discuss resonances in the
magnetoconductance due to the Berry phase (for vanishing Aharonov-Bohm
flux $\Phi$); then we discuss phase shifts in the Aharonov-Bohm oscillations
induced by the Berry phase.

We consider first the magnetoconductance as function of the Berry phase
in the absence of an Aharonov-Bohm flux, i.e. $\Phi=0$.
We make the realistic assumption
that the dephasing length independent of the tilt-angle 
can be made  to exceed $L$, say, $L_\phi=2.5 L$, giving
$\gamma=4.053\cdot 10^{-3}$ (this value for $\gamma$ is 100 times smaller than the one
chosen in Sec. \ref{exactsolution}).
In Fig.~1, we plot  the magnetoconductance
$\delta g$, Eq.~(\protect{\ref{moscexact}}),
as function of the tilt angle $\eta$ in the adiabatic regime, $\kappa=1$,
and find
pronounced resonance peaks at the Berry
phase values $\Phi^g=0,1$ --in very good qualitative agreement with the general
result $\delta g^{\lsgtag}$, 
given in Eq.~(\ref{moscLSG}), even if we simply choose
$\gamma'=\gamma$. For comparison, we also plot (see Fig.~1) the
magnetoconductance
$\delta g$ outside
the adiabatic regime, i.e., for $\kappa=0.01$, where the resonances are (nearly)
absent--
demonstrating that adiabaticity is needed for the emergence of the Berry phase.
We note that above choice for the adiabatic parameter (i.e., $\kappa_0=1$) 
corresponds
to $\omega_{B_0} \tau=(2\pi)^2 l^2/(L^2d)\gg l^2/L^2d$. 
In particular, if
we follow
van~Langen et al.\ and choose $L/l=500$ (i.e., a typical ratio for a mesoscopic
metal ring) we see that
$\kappa_0=1$ is equivalent to $\omega_{B_0} \tau=1.57\cdot 10^{-4}/d$. Note
that
we are orders of magnitude below the regime of Eq. (\ref{schrott}),
where $\omegabohr\tau\gg 1$.
Translated into  magnetic fields, $\kappa_0=1$ corresponds to
\begin{equation}
B_0={ 2(2\pi)^2\over gd}  {\vfermi l\hbar \over \mubohr L^2}=
{ 4\pi\over g \mubohr }{hD\over L^2}\, ,
\end{equation}
which, for $g=2$ and $d=3$~\cite{dimension}, gives
\begin{equation}
B_0=1.5 \times 10^{-6} {\vfermi l\over L^2} [Gs] =
4.5 \times 10^{-6} {D\over L^2} [Gs] \, .
\end{equation}
To illustrate this with concrete numbers we
assume the Fermi velocity $\vfermi=10^{6}\,{\rm ms}^{-1}$ and
the ring circumference $L=7\,\mu{\rm m}$, and again $L/l=500$.
We then find that the field corresponding to $\kappa=1$ is about 
$400\,{\rm G}$. The resonance structure due to the Berry phase
starts to emerge for $\kappa$ at around $0.1$, i.e., for fields 
of the order of $40{\rm G}$.  Finally, we note that when the tilt 
angle $\eta$ is varied, then typically there will be a concommittant 
change of the Aharonov-Bohm flux $\Phi$.  This flux, however, can be
easily compensated by applying a field perpendicular to the ring such that
$\Phi$ again becomes an integral multiple of the flux quantum.
Note that the maximal fields required for this compensation are about 
ten Gauss, or so, for a ring of $L=7\,\mu{\rm m}$. Thus, such fields 
would have a negligible effect on the inhomogeneous field required for 
adiabaticity, except if $\eta$ is very close to $\pi/2$.

A further experimentally interesting scenario is that of the {\it phase 
shift\/} in the Aharonov-Bohm oscillation induced by the Berry phase.
In particular, this effect is most pronounced for half-integral Berry phases,
$\Phi^g=\pm 1/2$ (i.e., $\eta=\pi/3$ or $2\pi/3$), for which we expect
[see Eqs.~(\ref{moscadiabatic}) and (\ref{moscLSG})]
to get a phase shift in the Aharonov-Bohm oscillation
of the magnetoconductance
by the flux value $1/2$ (i.e., by one quarter of
the flux quantum $h/e$). Note that
in this case the sign of the oscillation slope (e.g. at $\Phi=0$)
gets reversed with respect
to the
case without Berry phase. This sign-reversal is reminiscent of similar
effects induced by spin-orbit scattering~\cite{Aronov}; it
is actually not unexpected, as the Zeeman term induces an effective
spin-orbit coupling due to the inhomogeneity of the magnetic field
\cite{LGB1,LGlong}.
This phase shift is shown in Figs.~2a and 2b, 
which show $\delta g $ as function
of the Aharonov-Bohm flux $\Phi$ for Berry phases $\Phi^g=0$ and $1/2$,
both in the adiabatic limit (i.e., $\kappa=1$) 
and with the choice $\gamma=0.1$ (i.e., $L=2 L_\phi$).
For the sake of comparison, in Fig.~2d we also
show a non-adiabatic case, $\kappa=0.1$, for which
the phase shift is absent.
The phase shift remains discernible down to about $\kappa=0.7$ before 
disappearing.
The adiabatic limit is fully reached at about $\kappa=10$, by which 
not only the phase shift (which is the important feature)
but also the amplitude becomes identical to
$\delta g^{\adtag}$ given in Eq.~(\ref{moscadiabatic}). The  amplitude
at $\kappa=1$ increases  about by 20~percent upon increasing the field to
$\kappa=10$.

To obtain realistic estimates for some physical parameters
we now concentrate on a Au ring and use the material parameters recently 
determined by Mohanty et al.\cite{Mohanty} (see sample Au-$1$ in their 
Table~I). The relevant values are:
$D=9\times 10^{-3}\,{\rm m}^{2}{\rm s}^{-1}$ and 
$\tau^0_\phi=3.41\times 10^{-9}\,{\rm s}$ 
(at a temperature of $11\,{mK}$), which give 
for the dephasing length $L_\phi^0=\sqrt{D \tau^0_\phi}=5.54\,\mu{\rm m}$.
Thus, the above choice $L=2 L_\phi^0$ requires a ring of circumference 
$L=11\,\mu{\rm m}$.  In this case, the field corresponding to $\kappa_0=1$ 
becomes $B_0=335\,{\rm G}$, and the limiting case, $\kappa=0.7$, at which 
the phase shift emerges, corresponds to $B=235\,{\rm G}$~\cite{footnote4}.

Precisely the same phase shift occurs in $\delta g^{\lsgtag}$, 
Eq.~(\ref{moscLSG}), as shown in Fig.~2. To get roughly the same 
amplitudes as in $\delta g$ we must account for the $\eta$--dependent 
dephasing in  $\delta g^{\lsgtag} $. To this end we 
choose an effective
$\gamma'=\gamma=0.1$  (for $\eta=\pi/2$) and 
$\gamma'=5\gamma=0.5$ (for $\eta=\pi/3$.
This phenomenological choice is not vital for
the qualitative behavior of $\delta g^{\lsgtag} $,
but it does allow us to estimate an effective dephasing length $L'_\phi$, 
as we now explain.
First we note that the (peak-to-peak) amplitude of the magnetoconductance
$\delta g$
for $\Phi^g=1/2$
is considerably
reduced (by about a factor of
$25$) with respect to that for $\Phi^g=0$. As is clear by now, this is
due to the $\eta$--dependent
dephasing terms.
Now, without such dephasing
the Aharonov-Bohm amplitudes for  $\Phi^g=0$ and $\Phi^g=1/2$ would be equal
[see, e.g., Eq.~(\ref{moscLSG}) with a  $\gamma'$ that is $\eta$-independent].
 Thus, the reduction of
the Aharonov-Bohm amplitude at $\eta=\pi/3$ (relative to that at $\eta=0$)
serves as a quantitative measure of the $\eta$-dependent
dephasing. Expressed in terms of
an effective dephasing length, $L'_\phi=L/2\pi \sqrt{\gamma'}$,
we find $L'_\phi=2.5\mu$m, for the particular values chosen above 
(i.e., $\gamma'=0.5$, and $L=11\,\mu{\rm m}$). This dephasing length should
be compared with above value $L_\phi=L/2 =5.5\,\mu{\rm m}$
(corresponding to $\gamma=0.1$ and $L=11\,\mu{\rm m}$).

Finally, there is also the
 usual (spin-independent) dephasing arising from the field $B_z$ penetrating
a ring of finite width $a$. On the one hand, we need a sufficiently large
field so as to reach adiabaticity, and on the other hand such a field can
induce dephasing. Thus, to satisfy these conflicting requirements in an
optimal way
we should consider rings with a width $a$ as small as possible.
To get a rough estimate for such a width,  we take for the field
$B_z=B\cos\eta$ and insert this into the standard formula~\cite{Aronov},
$L_\phi^{B_z}=\sqrt{3} \phi_0/2\pi a B_z$. We now require 
that this dephasing length
should not become (much) smaller than $L_\phi^0$, so we choose
$L_\phi^{B_z}=L_\phi^0=5.5\,\mu{\rm m}$.
On the other hand, the field required for adiabaticity is about 
$B=200\,{\rm G}$,
and together with $L_\phi^{B_z}=5.5\,\mu{\rm m}$ and $\eta=\pi/3$
this corresponds to a ring width $a$ of the order of $20\,{\rm nm}$.
Note that as the effective dephasing length is obtained via
$1/(L^0_\phi)^2 +1/(L^{B_z}_\phi)^2$, the dephasing
effect due to $B_z$ penetrating the sample
increases $\gamma$ by a factor of two (i.e., $\gamma=0.2$).
As is seen from Fig.~2, the cases $\gamma=0.1$ and $\gamma=0.2$ behave in
the same way, i.e., 
with phase shift,
but the amplitude of $\delta g$ for $\eta=\pi/3$
and $\gamma=0.2$
is now reduced by a factor of $52$ compared with $\delta g$ for $\eta=0$
and $\gamma=0.1$. (Note that for $\eta=0$ the magnetic field for the
Aharonov-Bohm
oscillations can be chosen to be very small, so that $L^0_\phi$ dominates
over $L^{B_z}_\phi$ and thus $\gamma=0.1$.)\thinspace\ 
Finally, we note that the field component $B_z=B\cos\eta$ gives rise to an
Aharonov-Bohm phase $\Phi_z=L^2 B_z/4\pi$ that is, in general, not equal to
$n\phi_0$ (with $n$ integral). Therefore, this offset flux $\Phi_z$ must be
accounted for in order to assign the above phase shift unambiguously to the
Berry phase $\Phi^g=1/2$. For instance, for $L=11\,\mu{\rm m}$, we need $B_z^0=4.2\,{\rm G}$ in order to generate one flux quantum $\phi_0=h/e$ 
through the ring.  Now consider $\eta=\pi/3$, and, say, $B=200\,{\rm G}$, 
i.e., $B_z=100\,{\rm G}$.  To compensate the off-set $\Phi_z$, we need to 
increase $B_z$ by, say, $5\,{\rm G}$ to $B_z=105\,{\rm G}$, in which case $B_z/B_z^0=\Phi_z/\phi_0$ becomes an integer ($=25$).

The amplitude-reduction mentioned above  
demands 
sufficient experimental resolution, which we now estimate.
For the parameter values given above for an Au ring
and for $\eta=\pi/3$, we find (cf.~Fig.~2c) that
the peak-to-peak amplitude of $\delta g$ is about
$5.3\times 10^{-3}\times (e^2/\pi\hbar)$ for an effective $\gamma=0.2$.
The relative ratio,
$\delta g/g\propto \delta R/R$,
thus becomes of the order of $10^{-4}$ for a ring resistance 
$R\propto 1/g$ of the order
of $30\cdot (L/\mu$m) Ohms~\cite{Mohanty}, and $L=11 \mu$m.
Such sensitivity, as well as all the parameters estimated above, 
appear to be within present-day experimental reach.
Further scenarios for the Berry phases in transport  can be easily
worked out (see also Ref.~\onlinecite{LSG}).

It should be obvious by now that the  explicit agreement between 
$\delta g$ in the adiabatic limit and $\delta g^{\lsgtag}$
unambiguously demonstrates (and reinforces the general points
made in the previous subsections) that the adiabaticity criterion, 
Eq.~(\ref{adiabaticity}), is sufficient for the existence of the 
Berry phase and that, moreover, there exist physical regimes
where this Berry phase can be observed in magnetoconductance
oscillations (and other quantities).
By contrast, the far more stringent criterion Eq. (\ref{schrott}) 
is certainly not necessary, 
and therefore sets unwarranted demands on experimental searches for 
Berry phase effects. 

\subsection{Physical Interpretation of the Dephasing Terms}
\label{physical}

We now briefly return to the issue of the source of dephasing in the 
Hamiltonian $h(j)$ given in Eq.~(\ref{hfinal}), as well as 
its physical interpretation.
For this purpose we assume from the outset that we are in the adiabatic
regime, $\kappa\gg 1/(2\pi)^2$, and simply retain the leading contributions
when finding the inverse of $\gamma-h(j)$.
This allows us to identify those terms in the Hamiltonian $h$ that
are responsible for the dephasing.

{}From the matrix representation~(\ref{matrixh}) of $h(j)$ it is 
straightforward to see that only those matrix elements are important 
in the adiabatic limit that are 
simultaneously either diagonal or off-diagonal in both spin subspaces. 
No other matrix elements contribute at the leading order, $\kappa^2$, for 
the determinant or sub-determinants of $\gamma-h(j)$ that are necessary to
calculate the inverse. Thus we can replace $h(j)$ by the matrix
\begin{equation}
-\pmatrix{
(j-f\cos\eta)^2 + a  &   0
& 0       &   a   \cr
0   &  j^2 +a -i 2\kappa
 &   0   &   0     \cr
0     &   0
& j^2 +a +i 2\kappa
&  0    \cr
a   &    0
&    0
 & (j+f\cos\eta)^2  +a \cr
 }, 
\label{matrixh.adia}
\end{equation}
and we see that it is only the term ${f^2\over 2}
(1+\sigma_{1x}\sigma_{2x})\sin^2{\eta}$ in $h(j)$
that causes dephasing and leads to those $\eta$-dependent terms
in $\delta g^{\adtag}$ that are absent in $\delta g^{\lsgtag}$ (apart from
the differences in $\gamma$ and $\gamma'$).
Now, the first term, ${f^2\over 2} \sin^2{\eta}$, has already been identified
in the discussion of the exact solution (for $f=1$) for a
propagator containing only a single spin-1/2~\cite{LGlong}.
In a general path-integral approach,
this term has been interpreted as a consequence of quantum fluctuations:
The particle trajectory fluctuates around its classical path and
these fluctuations in turn lead to a fluctuating local magnetic field. Such
fluctuations, however,
violate the  standard assumption
underlying the  adiabatic approximation that the field should vary smoothly
as a function of its parameters (in the present case
the parameter is given by the position $x(t)$ of the particle on the
ring). We have pointed out previously (see Sec.~VI~F in 
Ref.~\onlinecite{LGlong})
that this term might lead to deviations from the adiabatic approximation, 
which is valid only for smooth variations.

The second term,
${f^2\over 2} \sigma_{1x}\sigma_{2x}\sin^2{\eta}$, is new, and
describes an effective
spin-spin interaction induced by the inhomogeneity of the magnetic field  
(i.e., in the Cooperon, the path and its time-reversed partner
are interacting with each
other via their respective spins). This interaction between spin 1 and spin 2
is transmitted via the orbital motion, and in this sense involves
a back-reaction of the Zeeman term on the orbital motion. However, as pointed
out in Sec.~\ref{semiclassical},
such back
reactions that act to suppress the phase-coherence
are consistently assumed to be negligible
in our semiclassical treatment. Thus, in Ref.~\onlinecite{LSG} we have
performed the adiabatic approximation  on the
propagators for the path and for  its time-reversed partner
separately and independently, and all possible dephasing effects are
included phenomenologically in terms of $\gamma'$ at the end.
This finally explains the apparent discrepancy between $\delta g^{\lsgtag}$ and
$\delta g^{\adtag}$. 
However, as shown in previous sections, this discrepancy
vanishes when allowing for $\eta$-dependent dephasing terms $\gamma'$ in
$\delta g^{\lsgtag}$.

\section{Conclusion}

By using the exact solution for the Cooperon we have shown that the
Berry phase leads to observable effects
in the magnetoconductance oscillation within the adiabatic regime
defined by Eq.~(\ref{adiabaticity}). This is in full agreement with
our previous findings~\cite{LSG}, and in contrast
to the claim made by van~Langen et al.~\cite{langen}. 
We have pointed out
the role of dephasing and emphasized its angle- and winding-dependence.
We have illustrated the general discussion with explicit examples which
reinforce
our optimistic outlook for the experimental search
of the Berry phase in diffusive
metallic samples.

\acknowledgments
We wish to acknowledge discussions with 
S.~van~Langen and Y.~Lyanda-Geller.
This work was supported by 
the Swiss National Science Foundation (DL and HS), 
the Deutsche Forschungsgemeinschaft as part of SFB 195 (HS), 
and by the U.S.~Department of Energy, Division of Materials Science, 
under Award No.~DEFG02-96ER45439 through the 
University of Illinois Materials Research Laboratory (PG).


\begin{figure}

\ifnum\draftfig=1
  \vspace*{0cm}
\else
\fi

\caption{ The (dimensionless) magnetoconductance 
$\delta g/(-e^2/\pi\hbar)$,
Eq.~(\protect{\ref{moscexact}}), 
as function of the tilt angle $0\leq \eta \leq \pi$.
Figure~1a shows $\delta g$ in the adiabatic limit 
(i.e., $\kappa =1$); 
Fig.~1b shows $\delta g$ outside the adiabatic limit 
(i.e., $\kappa =0.01$),
with a strongly reduced amplitude.
The remaining parameter values are
$f=1$, $\gamma=0.4053/100$ and $\Phi=0$.
Figure~1c shows the adiabatic result
$\delta g^{\lsgtag}/(-e^2/\pi\hbar)$, 
Eq.~(\protect{\ref{moscLSG}}), 
as function of tilt angle $0\leq \eta \leq \pi$, 
with $f=1$, $\gamma'=0.4053/100$ and $\Phi=0$. 
Note that Figs.~1a and 1c agree very well, qualitatively, and show 
pronounced resonances at integral values of the Berry phase 
$\Phi^g=0, 1,\ldots$. }
\label{Fig1}

\end{figure}

\begin{figure}

\ifnum\draftfig=1
  \vspace*{0cm}
\else
\fi

\caption{The (dimensionless) magnetoconductance 
$\delta g/(-e^2/\pi\hbar)$,
Eq.~(\protect{\ref{moscexact}}), 
as function of Aharonov-Bohm
flux $0\leq \Phi=2\phi/\phi_0 \leq 1$.
Figure~2a shows $\delta g$ at vanishing Berry phase in the 
adiabatic regime (i.e., $\eta=\pi/2$) and with parameter
values $\kappa=1$, $\gamma=0.1$; 
Fig.~2b shows $10 \delta g$ with Berry phase 1/2 in the adiabatic 
regime, i.e., $\eta=\pi/3$ and $f=1$, $\kappa=1$, and $\gamma=0.1$.
Note the phase shift (due to the Berry phase) by the amount $\Phi=1/2$ 
between Figs.~2a and 2b.
Figure~2c shows the same as Fig.~2b, except that here $\gamma=0.2$ 
(this accounts for the dephasing due to $B_z$, see text).
Figure~2d shows $\delta g/(-e^2/\pi\hbar)$ as function of Aharonov-Bohm
flux $\Phi=2\phi/\phi_0$, but outside the adiabatic regime:
$10 \cdot \delta g$ at Berry phase $\Phi^g =1/2$, 
i.e., $\eta=\pi/3$ and $f=1$, and $\kappa=0.1$, and $\gamma=0.1$. 
Note that there is no phase shift, which shows that the Berry phase 
is not yet in effect.
Figures~2e and 2f show the magnetoconductance in the adiabatic limit,
$\delta g^{\lsgtag}/(-e^2/\pi\hbar)$, Eq.~(\protect{\ref{moscLSG}}),
as function of Aharonov-Bohm flux $\Phi=2\phi/\phi_0$.
Figure~2e shows $\delta g^{\lsgtag}$ with vanishing Berry phase, 
i.e., $\eta=\pi/2$, and $\gamma'=0.1$; 
Fig.~2f shows $10\cdot \delta g^{\lsgtag}$ with Berry phase $1/2$, 
i.e., $\eta=\pi/3$ and  $f=1$, and $\gamma'=5\cdot 0.1$ (the increased 
$\gamma'$ accounts for the $\eta$-dependent dephasing, see text).
Again there is a phase shift by $\Phi=1/2$ between Figs.~2e and 2f, 
in full agreement with the adiabatic limit of $\delta g$ as shown in 
Figs.~2a and 2b. }
\label{Fig2}

\end{figure}


\begin{references}
\bibitem{footnote0}
Strictly speaking, we should also allow for the possibility that higher 
orbital momentum states $j$ will contribute to adiabatic behavior (on the 
scale given by the Thouless energy) by multiplying the right-hand side of
Eq.~(\ref{adiabaticity}) by some momentum average $\langle j\rangle$ (see 
the analogous discussion in~\cite{LGlong}). Note that $\langle j\rangle$ 
can be quite complicated, as it  depends in principle on all parameters 
occurring in the Hamiltonian describing the physical situation at hand. 
However, typically it turns out that only the lowest few momentum states 
contribute to physical quantities (provided the damping is sufficiently 
small); thus, we will ignore such refinements here and simply put 
$\langle j\rangle=1$. We have checked the validity of this approximation
(i.e. $\langle j\rangle=1$)
with the exact solution 
for the magnetoconductance by testing out when 
the adiabatic limit is fully reached in the various parameter regimes, 
and found it to be a reasonable approximation in the present situation 
(and for $\gamma$ not exceeding $0.5$).
\bibitem{footnote1}
Note that the complete Berry phase is usually defined as $f(1-\cos\eta)s$,
with $s$ the spin value and for constant tilt angle 
(cf.~Ref.~\protect{\onlinecite{LSG}}). 
For two spin-1/2 particles the total spin $s$ is
integral. The Berry phase 
is then equivalent to $\pm f(n-\cos\eta)$ (with $n=0,\pm 1,\pm 2,\ldots$)
[see, e.g., Eq.~(\ref{moscLSG})]
and we shall simply refer to $\Phi^g=f\cos\eta$ as the Berry phase.
In particular, $\Phi^g=0$ and $\Phi^g=1$ are equivalent in their
effect, but they belong to different field configurations.
\bibitem{footnote3} 
Note that the Aharonov-Bohm resonances in $\delta g$ are also suppressed 
by the $\eta$-dependent dephasing terms, in general, but more gradually 
than the Berry phase resonances, some of which 
disappear entirely
as soon as $f>1$. This is related to the fact that the Berry phase itself 
depends on $\eta$ whereas the Aharonov-Bohm phase does not.
\bibitem{footnote5}
The additional factor of $2$ in $\sin{2\eta}$ is 
intended to provide a phenomenological account of 
the smaller period with respect to the Berry phase in the Cooperon resulting
from the interference of the path with its time-reversed partner.
For instance, when $\eta=\pi/2$, one can easily convince oneself that
the fluctuations in  the individual Berry fluxes  
of spin 1 and spin 2 must exactly cancel each other as the trajectories
that the spins describe on the unit sphere in spin space are related
by time reversal symmetry. Thus, the dephasing must be reduced at $\eta=\pi/2$. 
While certainly plausible, this argument is only of suggestive value, 
as it ignores the effective interaction between the paths 
(see Sec.~\protect{\ref{physical}}).  Also, we ignore for the moment 
the fact that for $f>1$ such large $\gamma$'s actually violate
the self-consistency of the semiclassical approximation (see 
Sec.~\ref{semiclassical}).
\bibitem{footnote2} 
Note that if the  suppression of phase coherence is
only {\it partial\/}, the treatment is of course still self-consistent
within the semiclassical regime. This
is, e.g., the case for $f=1$ and $\eta=\pi/3$, for which the Berry phase 
is still visible albeit with a reduced amplitude
(see Sec.~\protect{\ref{observability}}).
\bibitem{dimension} 
We consider systems that are one-dimensional with respect
to the dephasing length $L_\phi$ but three-dimensional with respect to
the mean free path $l$.
\bibitem{footnote4}
We mention that the presence of magnetic impurities can help bring the 
adiabatic limit to lower fields. Without going into a detailed analysis, 
it seems plausible to assume that even for small fields (say, on the order of tens of
Gauss) the magnetic moments of typical impurities will align along the local
magnetic field direction (temperatures should be smaller than the Zeeman
splitting energy, typically in the sub-Kelvin range) after some relaxation 
time (typically on the order of $10^{-6}\,{\rm s}$). These magnetic 
impurities will produce short-ranged local fields that couples via the 
exchange interaction to the electrons passing by, and thus impart 
information about the field inhomogeneity on to the electron spin.
\end{references}
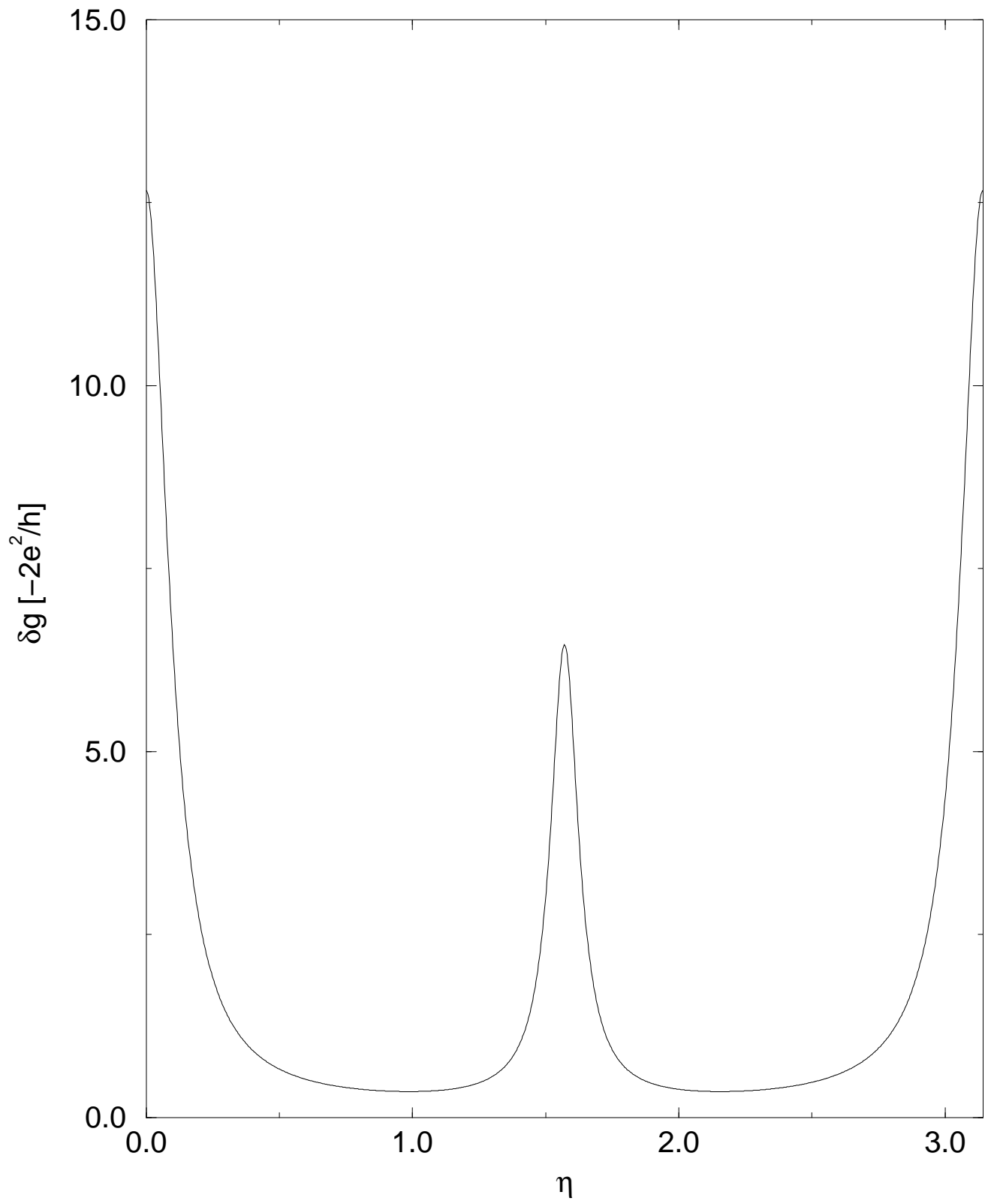
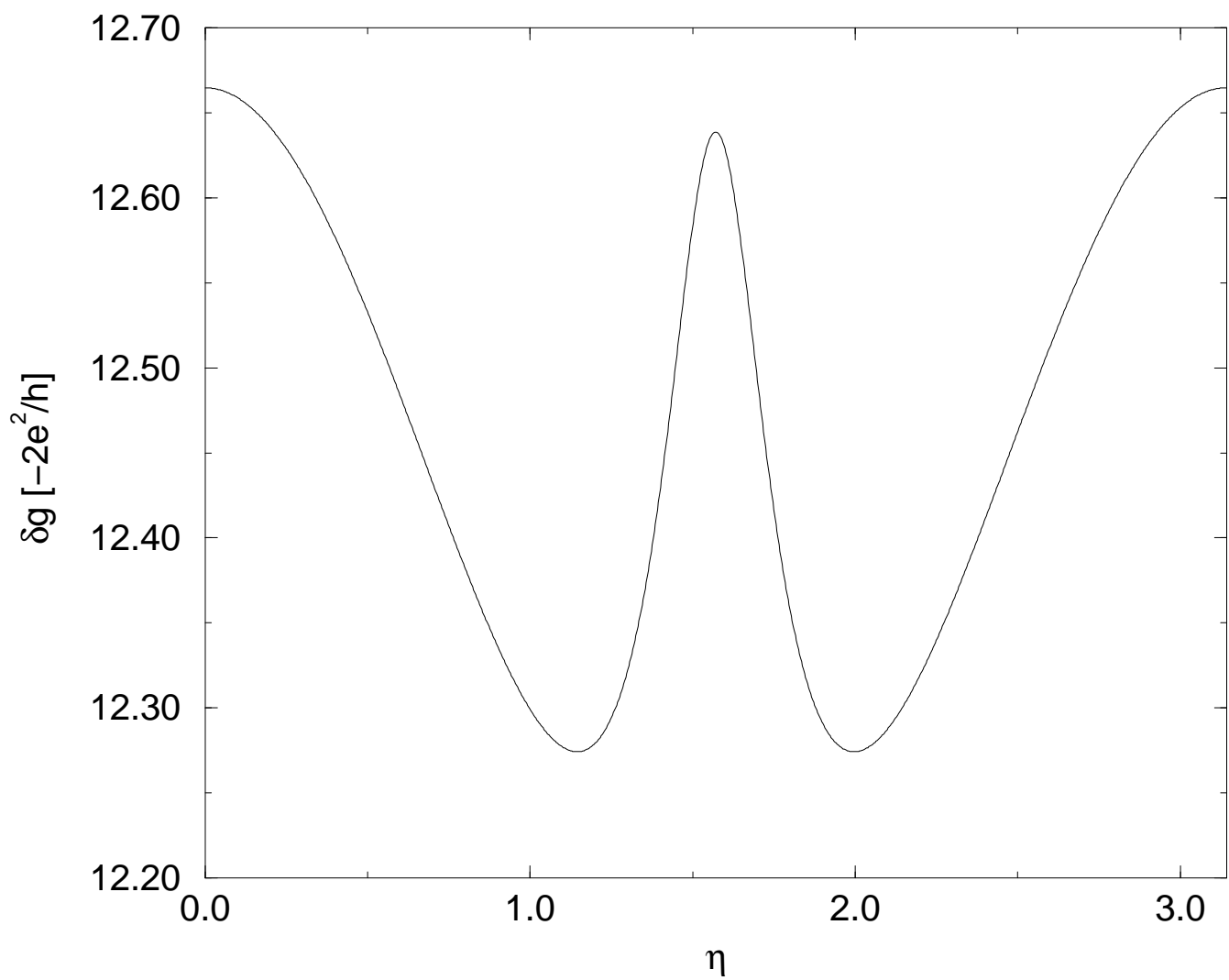
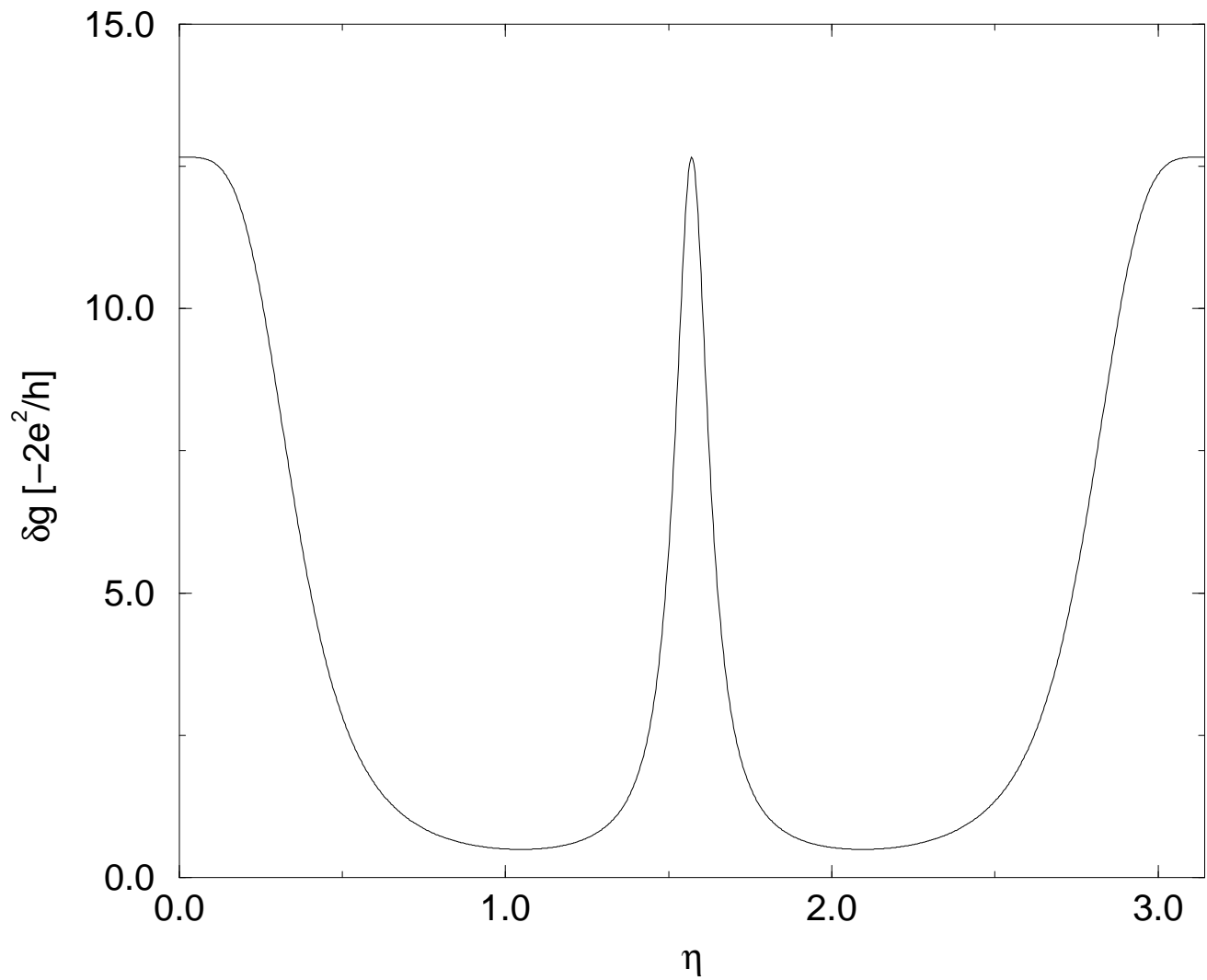
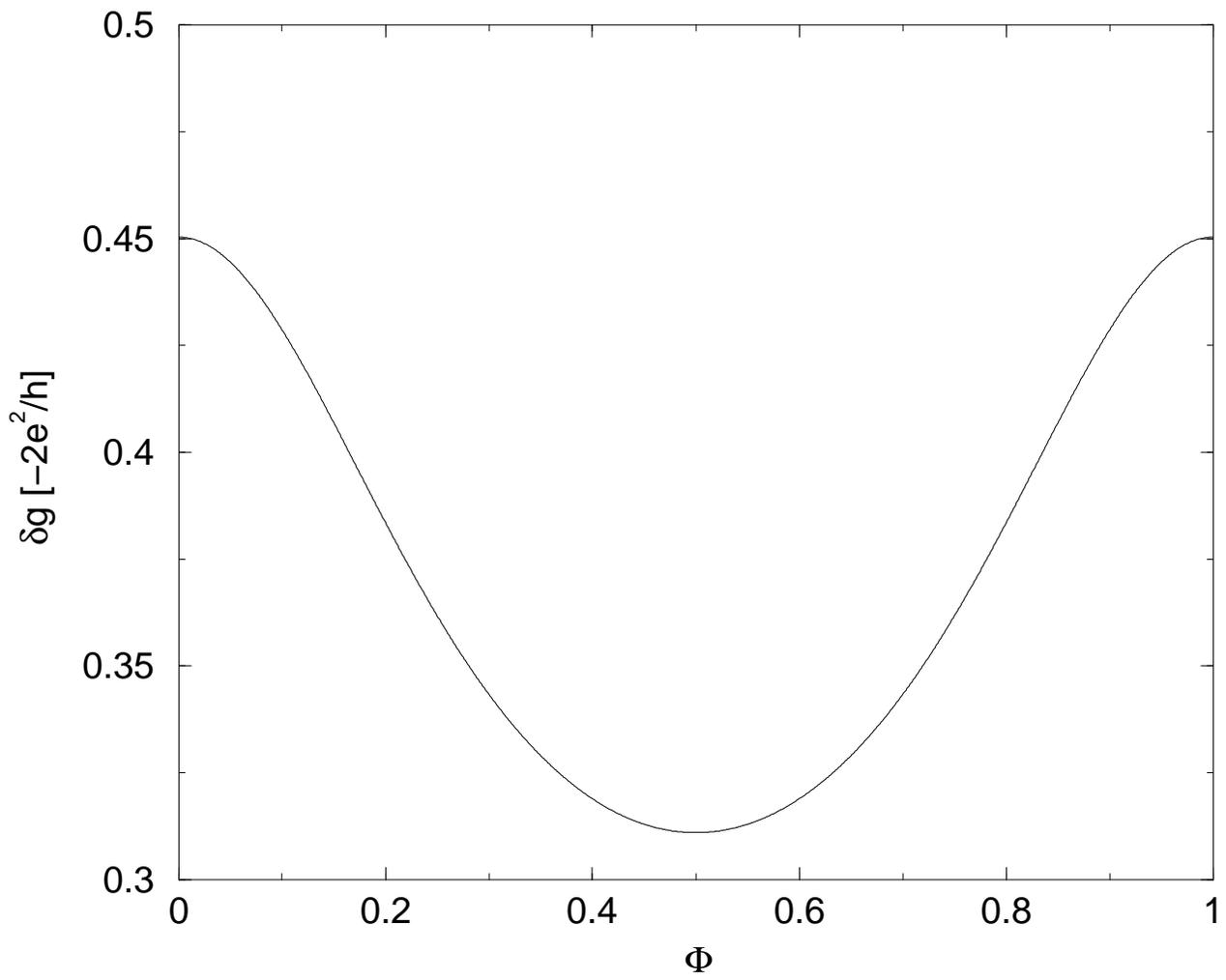
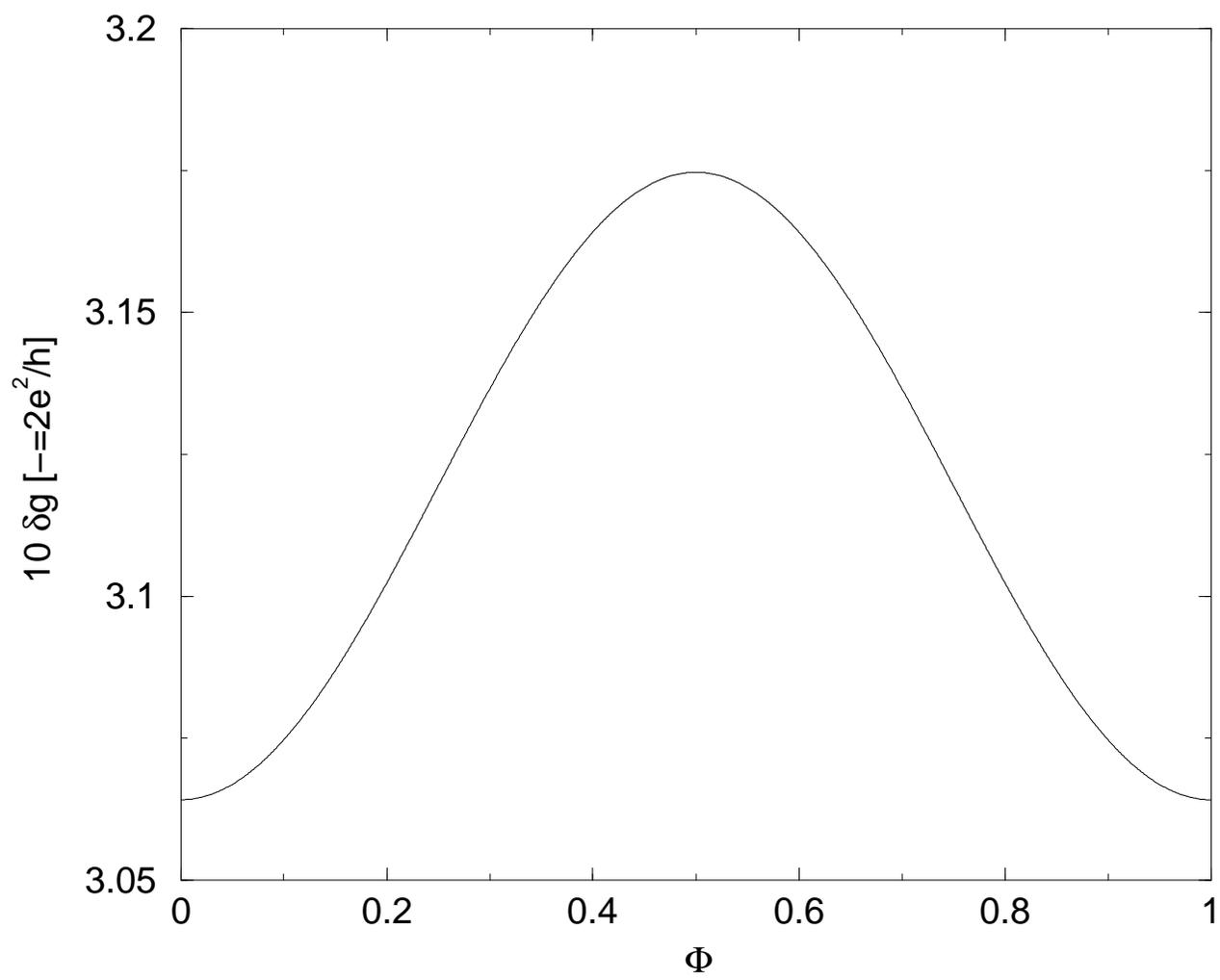
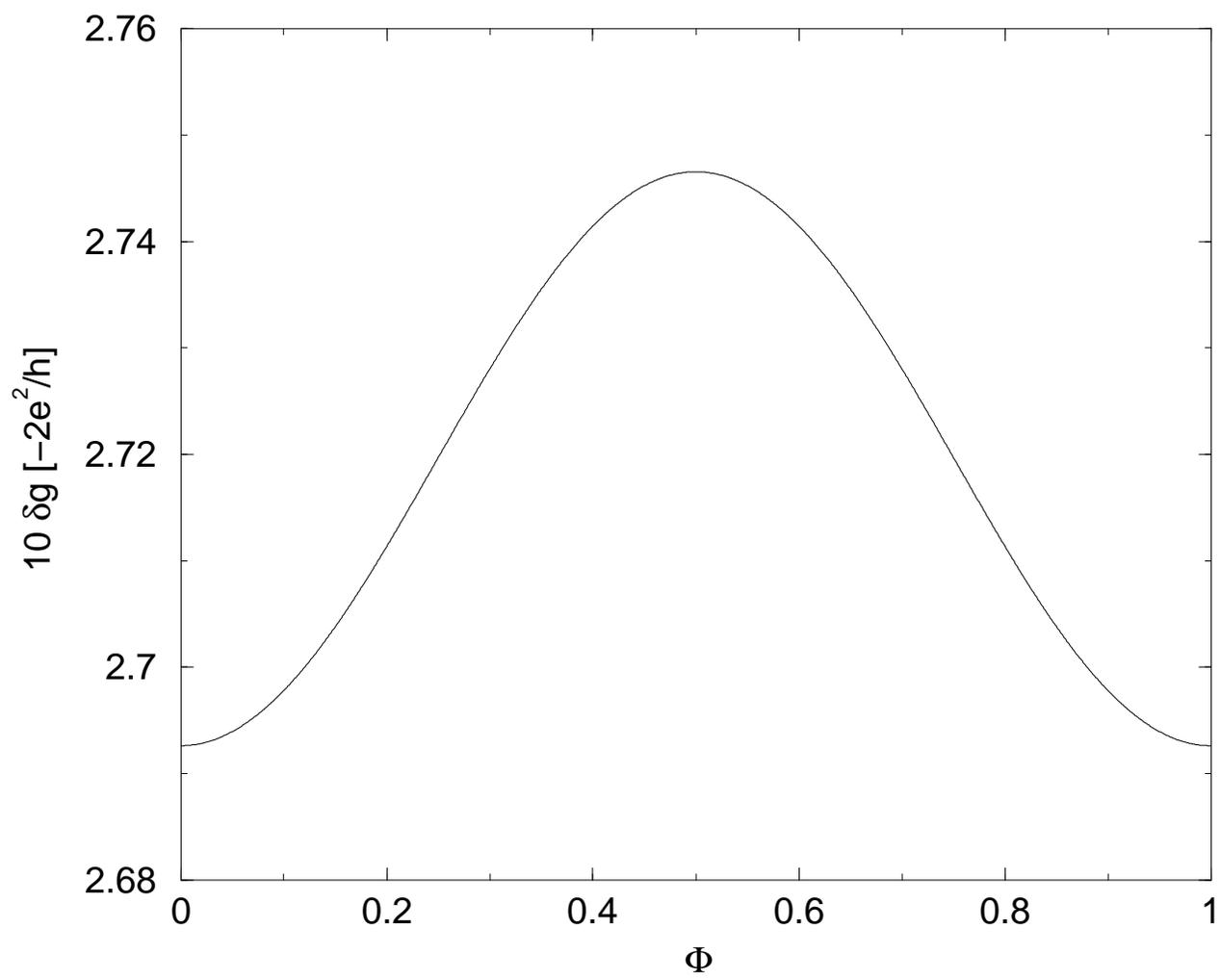
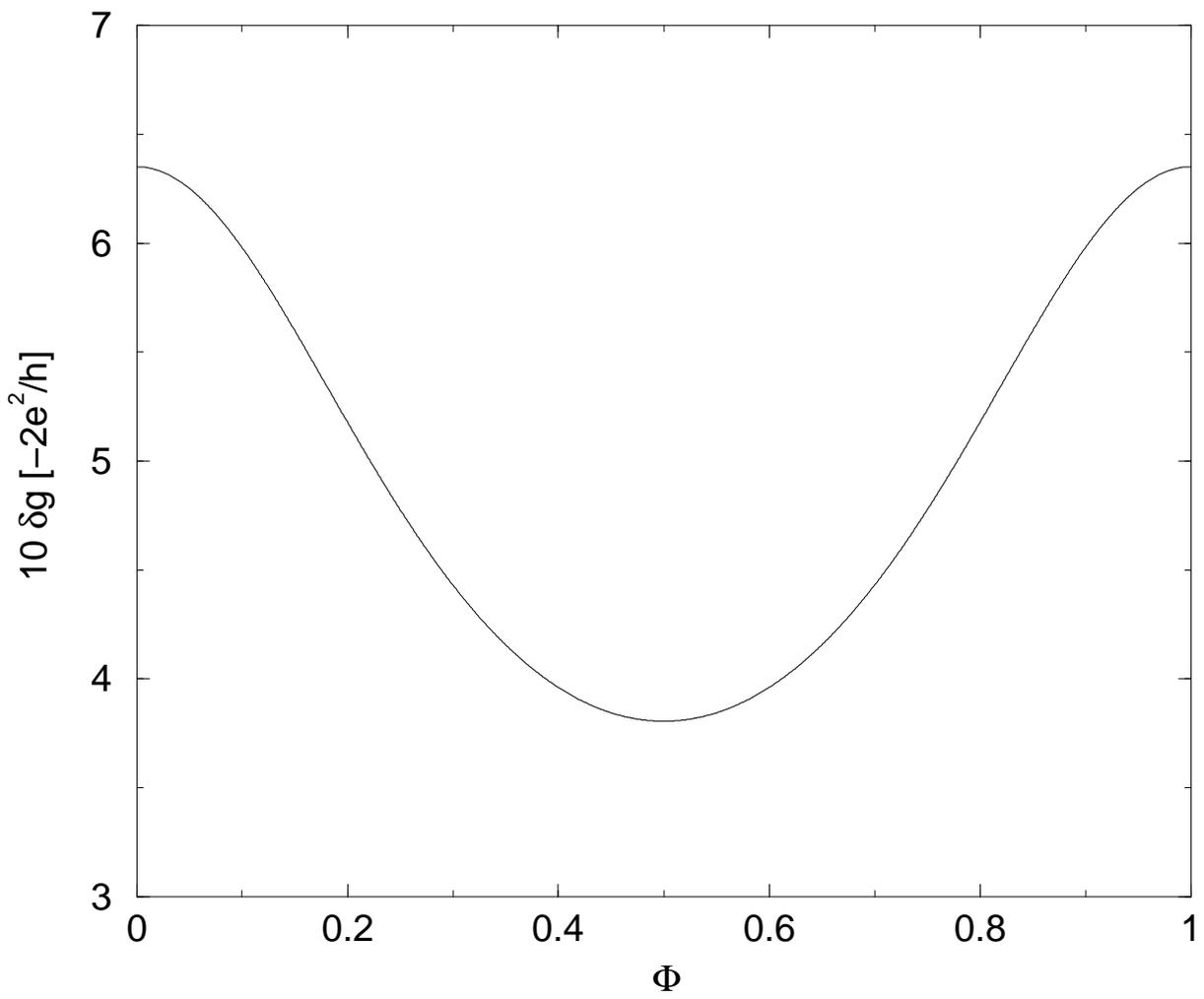
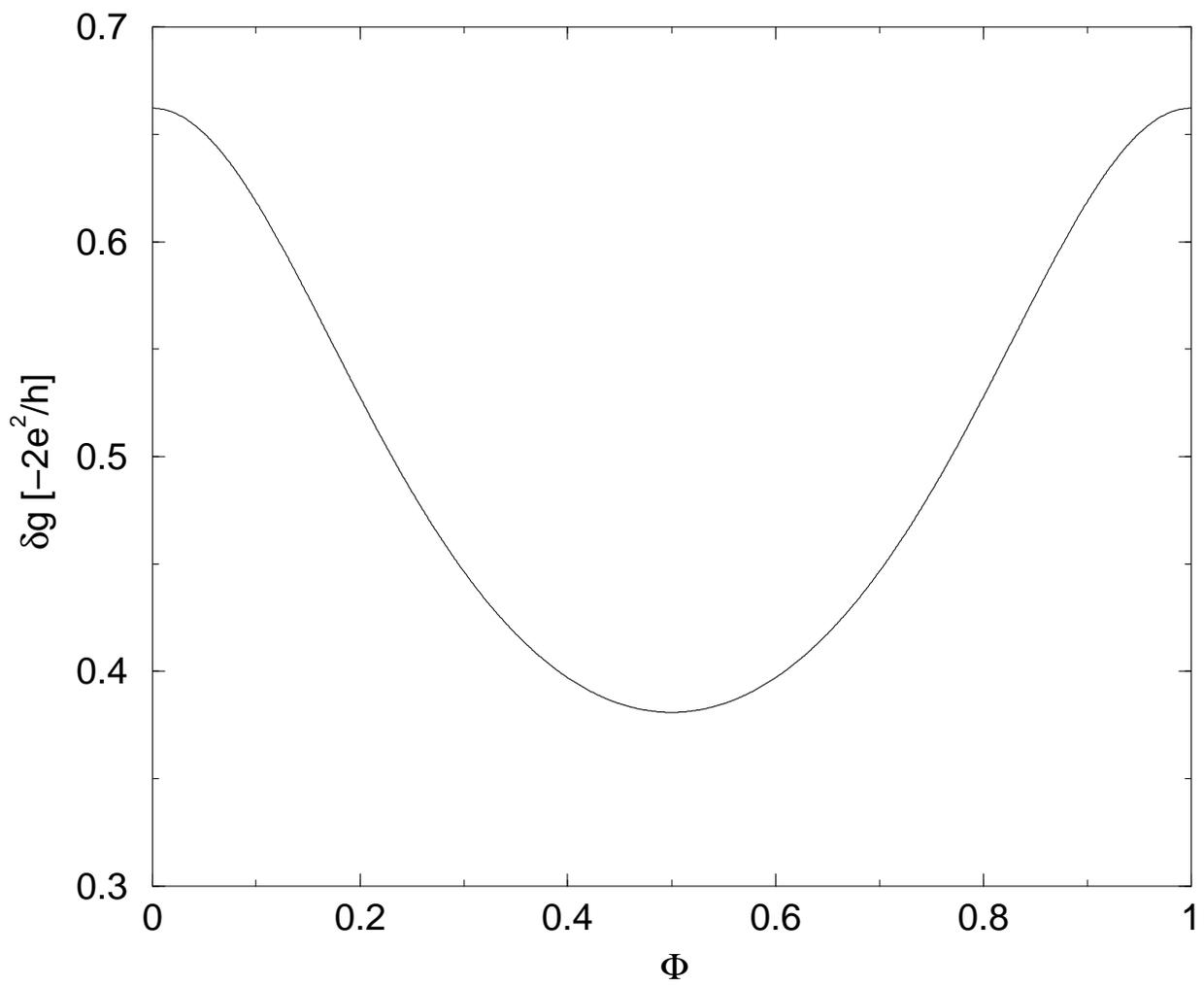
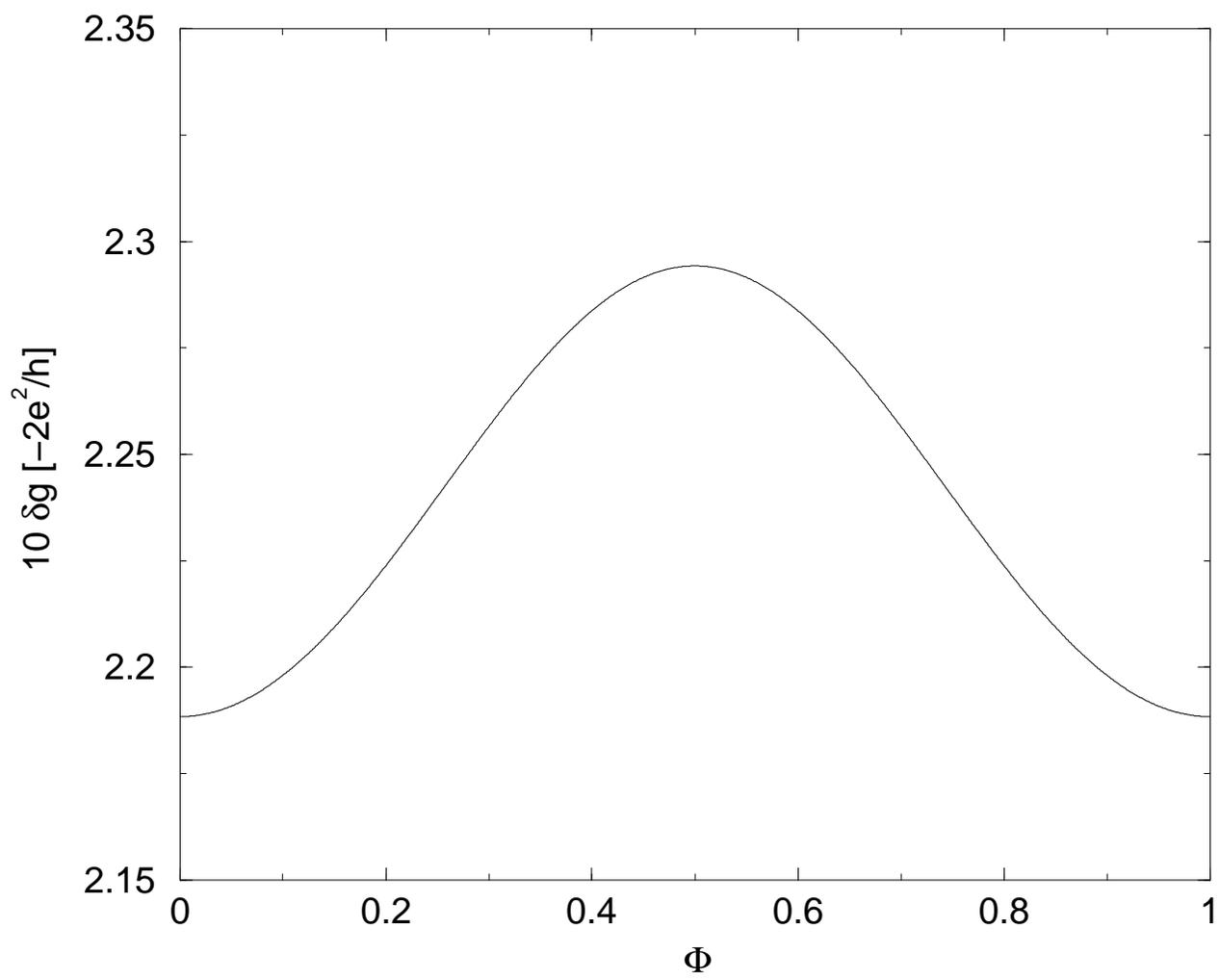
\end{document}